\documentclass[10pt,conference]{IEEEtran}
\IEEEoverridecommandlockouts
\usepackage{cite}
\usepackage{amsmath,amssymb,amsfonts}
\usepackage{graphicx}
\usepackage{textcomp}
\usepackage{algorithm}
\usepackage{algpseudocode} 
\usepackage{float}  
\usepackage{xcolor}
\usepackage{subcaption}
\usepackage{makecell} 

\def\BibTeX{{\rm B\kern-.05em{\sc i\kern-.025em b}\kern-.08em
    T\kern-.1667em\lower.7ex\hbox{E}\kern-.125emX}}

\begin{document}

\title{Performance Isolation for Inference Processes in Edge GPU Systems}

\author{\IEEEauthorblockN{1\textsuperscript{st} Juan José Martín }
\IEEEauthorblockA{\textit{DISCA} \\
\textit{Universitat Politècnica de València}\\
València, Spain \\
juamaros@upvnet.upv.es}
\and
\IEEEauthorblockN{2\textsuperscript{nd} José Flich}
\IEEEauthorblockA{\textit{DISCA} \\
\textit{Universitat Politècnica de València}\\
València, Spain \\
jflich@disca.upv.es}
\and
\IEEEauthorblockN{3\textsuperscript{rd} Carles Hernández}
\IEEEauthorblockA{\textit{DISCA} \\
\textit{Universitat Politècnica de València}\\
València, Spain \\
carherlu@upv.es}
}

\maketitle

\begin{abstract}
This work analyzes the main isolation mechanisms available in modern NVIDIA GPUs: MPS, MIG, and the recent Green Contexts, to ensure predictable inference time in safety-critical applications using deep learning models. The experimental methodology includes performance tests, evaluation of partitioning impact, and analysis of temporal isolation between processes, considering both the NVIDIA A100 and Jetson Orin platforms. It is observed that MIG provides a high level of isolation. At the same time, Green Contexts represent a promising alternative for edge devices by enabling fine-grained SM allocation with low overhead, albeit without memory isolation. The study also identifies current limitations and outlines potential research directions to improve temporal predictability in shared GPUs.
\end{abstract}

\begin{IEEEkeywords}
Deep learning models, Ensemble of neural networks, Process isolation, GPU, Multi-Process Service (MPS), Multi-Instance GPUs (MIG), CUDA, Inferences, Green Contexts (GC)
\end{IEEEkeywords}

\section{Introduction}
Artificial intelligence based on deep learning models (DLMs) is becoming increasingly common across various industries, including e-commerce, image and video analysis, and finance, among many others. However, in applications where malfunctions can result in the loss of human life or environmental damage, this technology is subject to strict compliance with certification standards (e.g., ISO 26262\cite{ISO26262} in the automotive domain). This article focuses on meeting the predictability requirements of tasks executed on a graphics processing unit (GPU) in the context of safety-related applications.

To enable the use of DLMs in the context of functional safety applications, ISO/IEC TR5469~\cite{TR5469} proposes using prediction strategies based on diverse neural network ensembles~\cite{BSC, DiversityAISafety}. This technique avoids relying on the output of a single neural network as the final prediction. Instead, the exact inference is performed across multiple models, and the final result is determined by a voting mechanism among all models. This approach enhances the overall robustness \cite{lakshminarayanan2017simple} and reliability \cite{sagi2018ensemble} of the system.

However, in high-criticality tasks, not only is the accuracy of the prediction important, but the timing guarantees provided by the system are also necessary. In other words, a DLM must produce a correct prediction and do so within the required time constraints. Meeting this requirement is particularly challenging when GPUs are concurrently executing multiple tasks.

With these two premises in mind, this study aims to analyze the alternatives modern GPUs offer to enable temporal isolation between multiple concurrently running tasks.

\section{Background}

Nowadays, numerous GPU-based systems exist for neural network model inference, ranging from large computing clusters to embedded systems with power consumption below 10 W \cite{ngo2025edge}, including general-purpose GPUs.

Despite this, in almost all cases, computational resources are underutilized, especially when performing inferences with small batch sizes, such as in real-time processing systems or autonomous driving platforms.

This underutilization opens the possibility of parallelizing inference processes on the GPU, a strategy widely employed on CPUs. This approach addresses two issues simultaneously: on one hand, it enables full utilization of available computational resources, and on the other, it facilitates the use of neural network ensemble strategies.

\begin{figure}[h]
\centering
\includegraphics[ width=0.45\textwidth]{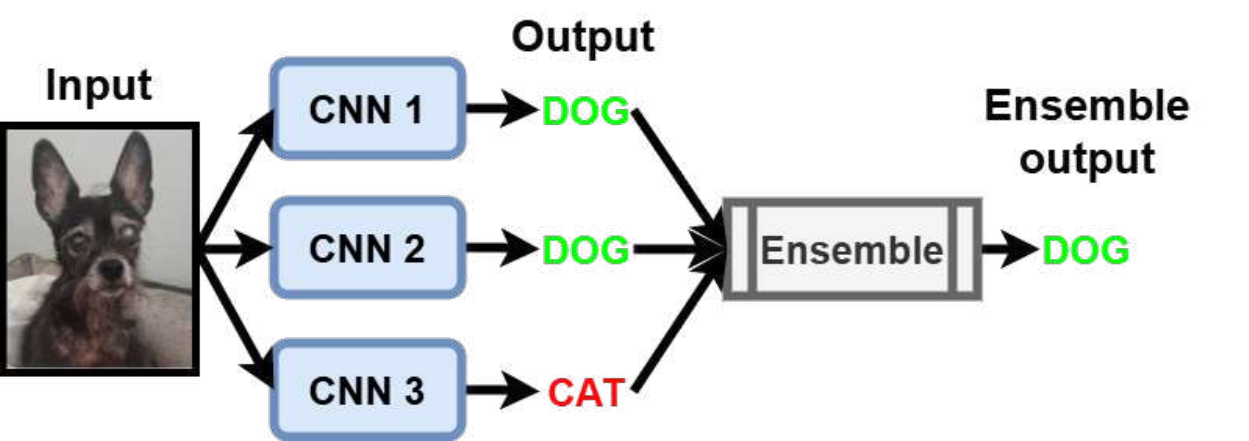}
\caption{Conceptual diagram of the operation of a neural network ensemble system.}
\label{ensemble}
\end{figure}

Neural network ensembles is a technique to improve system accuracy, where the prediction is not derived from a single model but from the combination of multiple models evaluated on the same input, as illustrated in Figure \ref{ensemble}. This is one of the most common strategies to meet the precision requirements demanded by high-criticality systems.

In many cases, meeting these standards requires duplicating or tripling computing hardware to perform inferences simultaneously. However, this article focuses on mechanisms provided by modern GPUs to perform parallel inferences, taking advantage of underutilized resources. We also analyze the temporal guarantees offered by each approach, aiming to propose solutions that allow multiple models to be inferred in parallel on a single GPU, efficiently exploiting resources and evaluating the impact on execution time. These range from lightweight alternatives that facilitate context switching between processes to strict partitioning and resource allocation strategies.

For this study, we focus primarily on two types of standardized hardware solutions:

\begin{itemize}
    \item General-purpose GPUs, represented by the NVIDIA A100, which feature compute units based on Streaming Multiprocessors (SMs) and SIMD-style vector processors.
    
    \item Edge computing-focused systems, represented by the Jetson Orin Nano and Orin AGX, based on unified System-on-Chip (SoC) platforms, which include a GPU also based on SMs and shared DRAM memory, allowing the CPU, GPU, and specialized accelerators to access the same memory space. These edge-oriented systems offer energy-efficient platforms that are ideal for performing real-time local inference.
\end{itemize}

Together, these two types of hardware enable the exploration of inference parallelization strategies, resource optimization, and process isolation evaluation, opening up new possibilities for critical real-time applications.

\section{Isolation mechanisms in modern GPUs}
Currently, few technologies are available to achieve parallel execution of processes on the GPU. The leading solutions are proprietary technologies developed by NVIDIA. Below, the most notable ones are listed and described along with their key features:

\subsection{Multi-Process Service}

Multi-Process Service (MPS) \cite{nvidia_mps} is a software-based solution designed to unify the individual CUDA contexts of multiple processes running on a single GPU. It represents NVIDIA's default mechanism for enabling concurrent execution, allowing true parallelism when sufficient GPU resources are available. In constrained resources, MPS improves performance by minimizing the overhead of context switching between processes.

\subsection{Multi-Instance GPUs}

Multi-Instance GPUs (MIG) \cite{nvidia_mig} is a hardware-based solution available on specific modern NVIDIA GPUs. This technology enables the creation of physically isolated partitions within a single GPU, both in compute units and memory resources.

The minimum compute unit that can be assigned to a MIG partition is a GPU Processing Cluster (GPC), which consists of a set of Texture Processing Clusters (TPCs). TPCs are modular processing blocks in NVIDIA GPU architectures, each typically containing two SMs, a texture unit, and a level-1 texture cache. This work uses an NVIDIA A100 GPU, which features 7 GPCs, each comprising 8 TPCs.

\begin{figure}[h]
\centering
\includegraphics[trim=50 417 50 54, clip, width=0.45\textwidth]{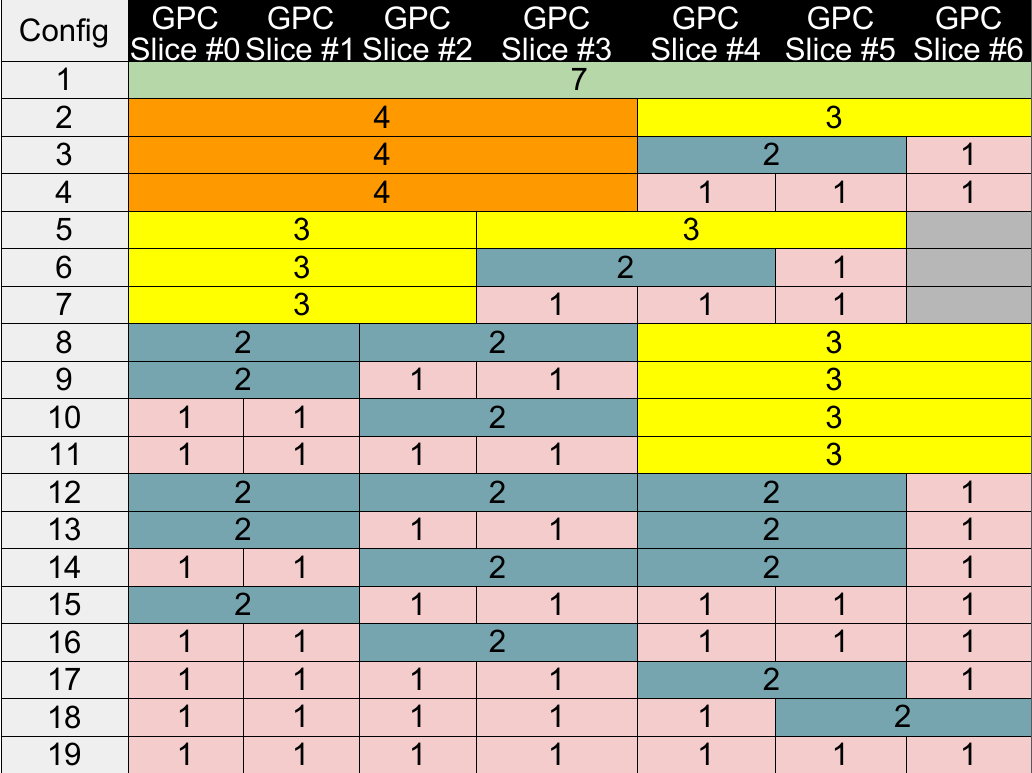}
\caption{Possible MIG instance configurations for the NVIDIA A100 GPU. Each color represents a possible MIG partition size based on the number of GPCs.}
\label{fig:MIG}
\end{figure}

Regarding the creation of MIG partitions, various configurations are available, depending on the number of GPCs allocated to each partition. These configurations are illustrated in Figure~\ref{fig:MIG}.

\subsection{Green Contexts}

Green Contexts (GC)\cite{nvidia_gc}, \cite{nvidia_gc_new} are a software-based solution introduced in CUDA version 12.4. This technology enables the creation of specialized CUDA contexts that limit the maximum number of Streaming Multiprocessors (SMs) accessible to any process executing within that context.

As with the previous technologies, this approach ensures the exclusive execution of the process on the assigned SMs; however, it does not provide isolation at the level of other hardware components, such as communication buses or memory subsystems.

However, Green Contexts allow for significantly more fine-grained resource allocation, treating individual SMs as the minimum unit of compute partitioning. According to the documentation, there is a minimum number of SMs per partition, as well as a step size by which the number of assignable SMs can increase. These constraints are imposed by the chip architecture, with the basic configurations summarized in Table \ref{tab:gc_sm_allocation}.

\begin{table}[H]
\centering
\caption{Minimum SM allocation and step size for Green Context partitions across different NVIDIA compute architectures.}
\label{tab:gc_sm_allocation}
\begin{tabular}{|c|c|c|}
\hline
\makecell{\textbf{Compute} \\ \textbf{Architecture}} & 
\makecell{\textbf{Minimum SMs} \\ \textbf{per Partition}} & 
\makecell{\textbf{SM Allocation} \\ \textbf{Step}} \\
\hline
6.X & 2 & 2 \\
7.X & 2 & 2 \\
8.X & 4 & 2 \\
9.0+ & 8 & 8 \\
\hline
\end{tabular}
\end{table}

\begin{figure}[h]
\centering
\includegraphics[ trim= 20 0 0 10, clip,width=0.50\textwidth]{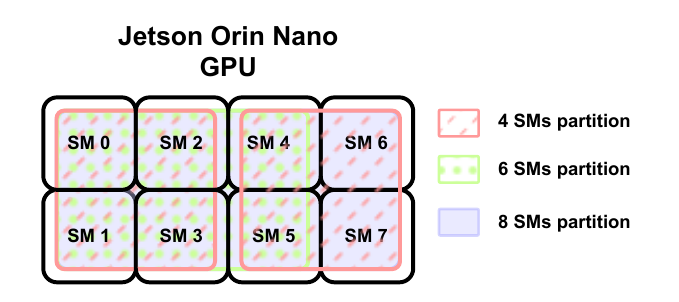}
\caption{Possible GC instance configurations for the NVIDIA Jetson Orin Nano GPU. Each color represents a possible GC partition size, based on the number of SMs and which SMs are utilized, depending on the assigned partition size.}
\label{fig_posi_GC}
\end{figure}

Similarly to MIG, both the maximum number of instances and the partition sizes are strongly tied not only to the architecture, but also to the amount of physical SM resources available on each GPU. In Figure \ref{fig_posi_GC}, we can see a graphical representation of the possible partitions that can be instantiated on the Jetson Orin Nano.

\section{Evaluation methodology}
\subsection{Experimental Setup}
\subsubsection{Hardware and software versions}
Given the nature of the experiments, it is necessary to employ different platforms on which the various technologies can be executed:

On the one hand, a general-purpose system equipped with an NVIDIA A100 GPU with 40 GB of memory, CUDA version 12.1, and PyTorch version 2.4.1, on which the experiments involving MPS and MIG technologies will be conducted.

On the other hand, as edge-oriented devices, the Jetson Orin Nano and Orin AGX development kits will be employed, utilizing the default software versions available in JetPack 6.2, on which experiments with MPS and Green Context technologies can be conducted.

\subsubsection{Models}

When selecting the models for the experiments, six of the most commonly used architectures in this type of study were chosen. These include classical convolutional neural networks, vision transformer (ViT) models, and more modern neural networks with increasingly complex topologies.

The weights, topologies, and biases were obtained from publicly available versions in the Hugging Face  repository \cite{HuggingFace}, all of which were pretrained for image classification on the ImageNet-1K dataset \cite{imagenet}. However, in this study, no metric related to the classification accuracy of the models will be considered.

For the NVIDIA A100, the pretrained models were utilized directly in their original format. In contrast, a distinct optimization process was employed for the Jetson Orin Nano. The models were converted into a highly optimized binary format known as a TensorRT engine file (.engine).

This optimization, performed through tools such as trtexec, is a standard practice in edge computing for several critical reasons. The process fuses layers and selects the most efficient GPU kernels, all tailored to the specific hardware. These adjustments enable a general optimization that significantly improves inference speed, reduces power consumption, and minimizes memory footprint, while maintaining competitive accuracy compared to other strategies \cite{opti} \cite{FP16}.

This approach ensures the experiments replicate real-world use-case scenarios where models must operate efficiently on resource-constrained devices, prioritizing performance metrics such as latency and throughput over the flexibility required during the training phase.

\subsection{Partitioning Performance Impact}

As with other studies that have already analysed similar technologies \cite{zhang2023migperfcomprehensivebenchmarkdeep}, it is important to understand the impact that these resource partitioning technologies can have on the regular operation of an inference process.

Therefore, before evaluating their capacity to isolate processes, the effects of MIG and GC partition sizes on the performance of an inference process were studied. Specifically, the effects of MIG were assessed on the A100, while those of GC were evaluated on the Jetson Orin.

The parameters to be analyzed are:
\begin{itemize}

\item \textbf{Throughput (Images/s)}: This indicates the number of inferences that a given network can perform per unit of time, in this case, inferences per second. For this test, inferences will be performed on a batch of 1.  

\item \textbf{Memory used (MiB)}: This parameter reflects the amount of GPU memory currently allocated and used by the active processes on that GPU.  

\item \textbf{Average power consumption (W)}: Represents the average of the instantaneous power measurements taken during the sampling intervals over the period in which the inference is being performed.

\end{itemize}

The experiments in this section are conducted using the ConvNeXt-Large \cite{convnet} neural network, which is the largest model employed in this work. For this evaluation, inference is performed with a batch size of 1 image.

\subsection{GPU Isolation Benchmark}

After becoming more familiar with these technologies, a series of experiments was designed to provide a clear understanding of the isolation capabilities of MPS, MIG, and GC, each within its specific platform. The goal is to assess whether these technologies are suitable for high-criticality areas and whether a parallel can be drawn between the behavior of MIG and GC, potentially using one as a substitute for the other in edge computing devices that currently do not support MIG.

To achieve this, the experiment is divided into two stages. First, the maximum guaranteed inference latency for each of the six models employed in this study, ConvNeXt Base \cite{convnet}, ConvNeXt Large, MobileNetV2 \cite{Mobil}, ResNet18 \cite{resnet}, ViT B 16 \cite{vit}, and ViT L 32 \cite{vit}, is determined.

To achieve this, an iterative process is performed in which each model exclusively executes inference on the GPU while progressively adjusting the time allocated for each inference. The algorithm begins by estimating an initial maximum inferences per second (IMS) value, calculated from the average of the five worst inference times from the first batch. The IMS is then incrementally increased until a timing violation occurs. This value is considered the maximum candidate, and a validation stage is performed using multiple batches to confirm stability. If the candidate frequency passes all validation tests, it is recorded as the maximum stable inference frequency; otherwise, the procedure is repeated with the subsequent lower frequency. The complete procedure is summarized in Algorithm \ref{alg:maxfreq}. For this study, the previously described models (M) were employed, with a number of inferences per batch of one thousand (N) and three validation batches (K).

\begin{algorithm}
\caption{Maximum Inference Frequency Search}
\label{alg:maxfreq}
\begin{algorithmic}[1]

\Require Model $M$, number of inferences per batch $N$, number of validation batches $K$
\Ensure Maximum stable inference frequency $f$

\Statex

\State \textbf{(Initial Estimation)}
\State $T \gets \textsc{RunInferenceBatch}(M, N)$
\State $W \gets \textsc{SelectWorstTimes}(T, 5)$
\State $f \gets 1 / \textsc{Average}(W)$

\Statex

\State \textbf{(Frequency Search)}
\State $stable \gets false$

\While{$stable = false$}

    \State $violations \gets \textsc{RunTimedBatch}(M, N, f)$

    \If{$violations = 0$}
        \State $f \gets \textsc{IncreaseFrequency}(f)$
    \Else
        \State $f \gets \textsc{DecreaseFrequency}(f)$
        
        \Statex
        \State \textbf{(Stability Verification)}
        \State $valid \gets true$

        \For{$i = 1 \to K$}
            \State $viol \gets \textsc{RunTimedBatch}(M, N, f)$
            \If{$viol > 0$}
                \State $valid \gets false$
                \State \textbf{break}
            \EndIf
        \EndFor

        \If{$valid = true$}
            \State $stable \gets true$
        \Else
            \State $f \gets \textsc{DecreaseFrequency}(f)$
        \EndIf

    \EndIf

\EndWhile

\State \Return $f$

\end{algorithmic}
\end{algorithm}

This maximum latency is used for tests conducted with MPS and the GPU stand-alone. In contrast, since MIG and GC are strict resource allocation technologies, a partition of 3 GPCs is used for MIG and 4 SMs for GC, as this is the maximum number of resources that can be allocated to generate two equal partitions. This allows obtaining the IMS value (\ref{eq}), which depends on the batch size and the maximum inference latency ($L_{\text{max}}$) and is later used for further tests.

\begin{equation}
\text{IMS} = \frac{\text{BATCH\_SIZE}}{L_{\text{max}}} = \frac{1}{L_{\text{max}}} \label{eq}
\end{equation}

Once the IMS value for each network and technology is determined, an experiment is designed to evaluate the isolation level provided by these technologies. The experiment involves two processes competing for GPU resources: one process of interest, which always performs inference at its maximum IMS, and a second process that gradually increases its IMS from 1 to its maximum. This setup enables the observation of how increasing the GPU workload affects the process of interest and provides a means to assess the capacity of each technology to maintain temporal guarantees independently.

A key aspect of this study is the concept of timeout, which occurs when an inference cannot be completed within the time allocated based on its IMS. Monitoring timeouts enables a quantitative assessment of whether the technologies under evaluation can ensure that the process of interest maintains its predefined inference time despite contention for GPU resources.

The experiments are conducted on two primary platforms. On the A100, we evaluate the standalone GPU, MPS, and MIG, using partition configuration 5 described in Figure \ref{fig:MIG}, which allocates two partitions with 3 GPCs each. On the Jetson Orin Nano, we also assess the standalone GPU and MPS, as well as GC, using the configuration of two GCs with 4 SMs each, as shown in Figure \ref{fig_posi_GC}. It is essential to note that the primary objective of this study is to assess whether these technologies can consistently provide temporal guaranties, rather than to compare absolute inference times across devices with substantially different computational resources.

\section{Experimental Results}
In this section, the results of the experiments presented in the previous section will be introduced and analyzed:

\subsection{Results of the partitioning performance impact employing MIG}

\begin{figure*}[htbp]
    \centering
    \begin{subfigure}[t]{0.25\textwidth}
        \centering
        \includegraphics[width=\textwidth]{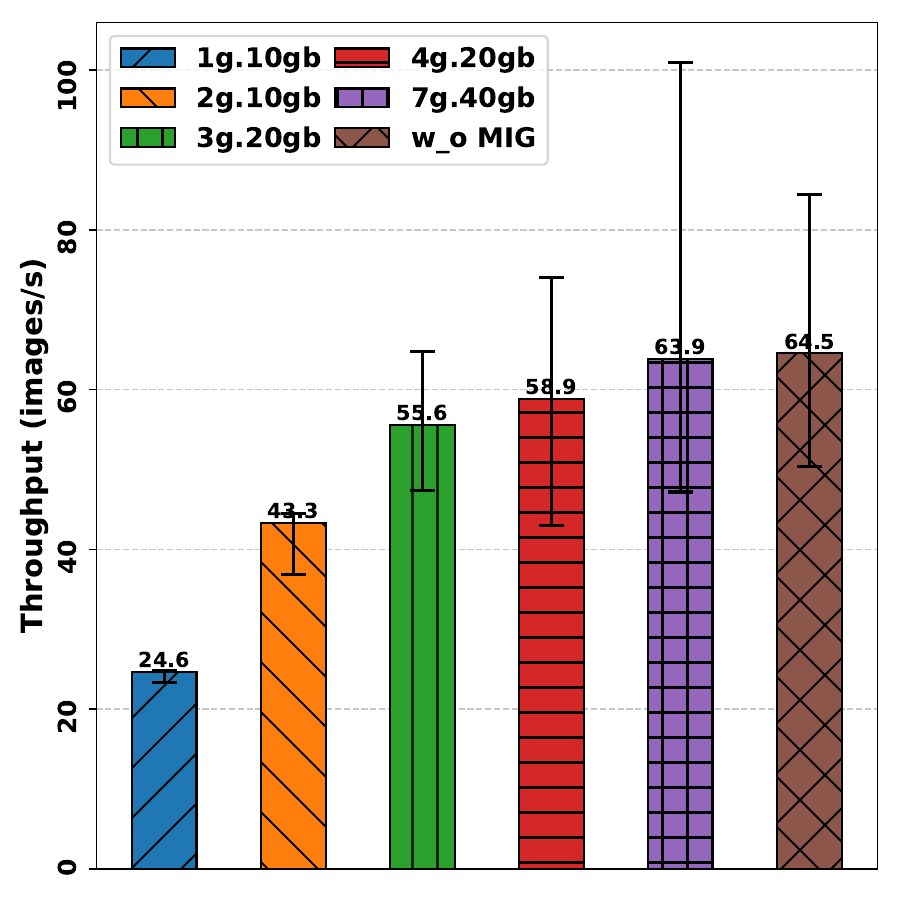}
        \caption{Throughput (images/s)}
        \label{fig:MIG_A}
    \end{subfigure}
    \hfill
    \begin{subfigure}[t]{0.25\textwidth}
        \centering
        \includegraphics[width=\textwidth]{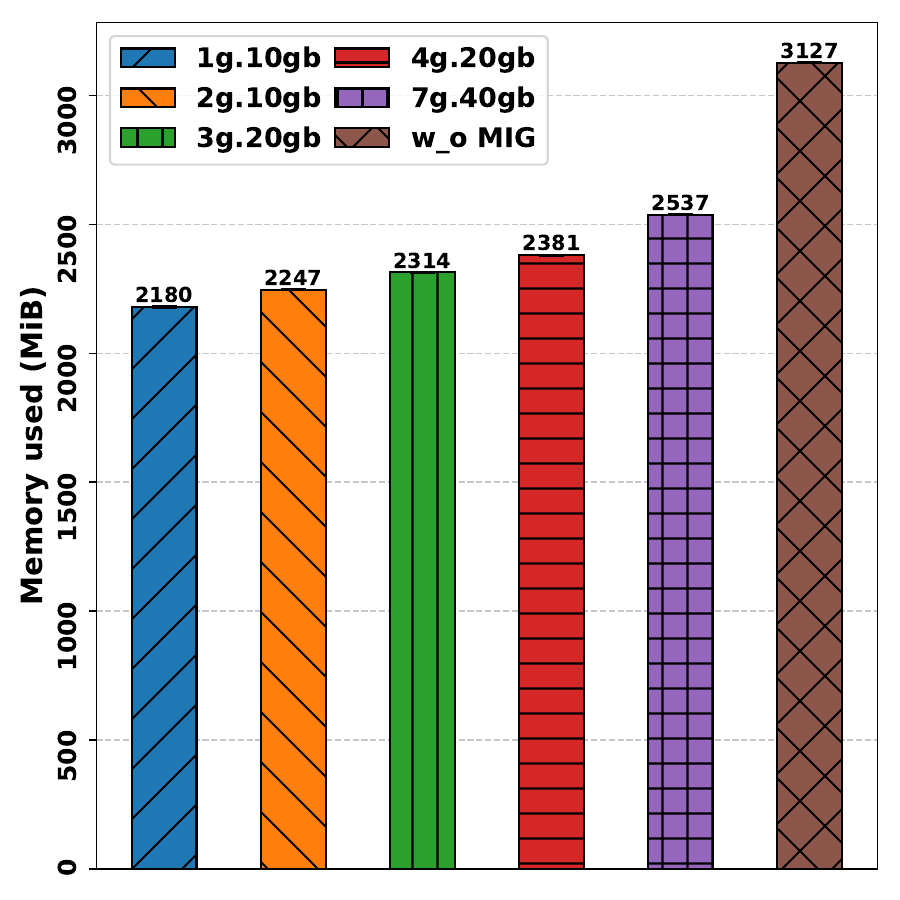}
        \caption{Memory used (MiB)}
        \label{fig:MIG_B}
    \end{subfigure}
    \hfill
    \begin{subfigure}[t]{0.25\textwidth}
        \centering
        \includegraphics[width=\textwidth]{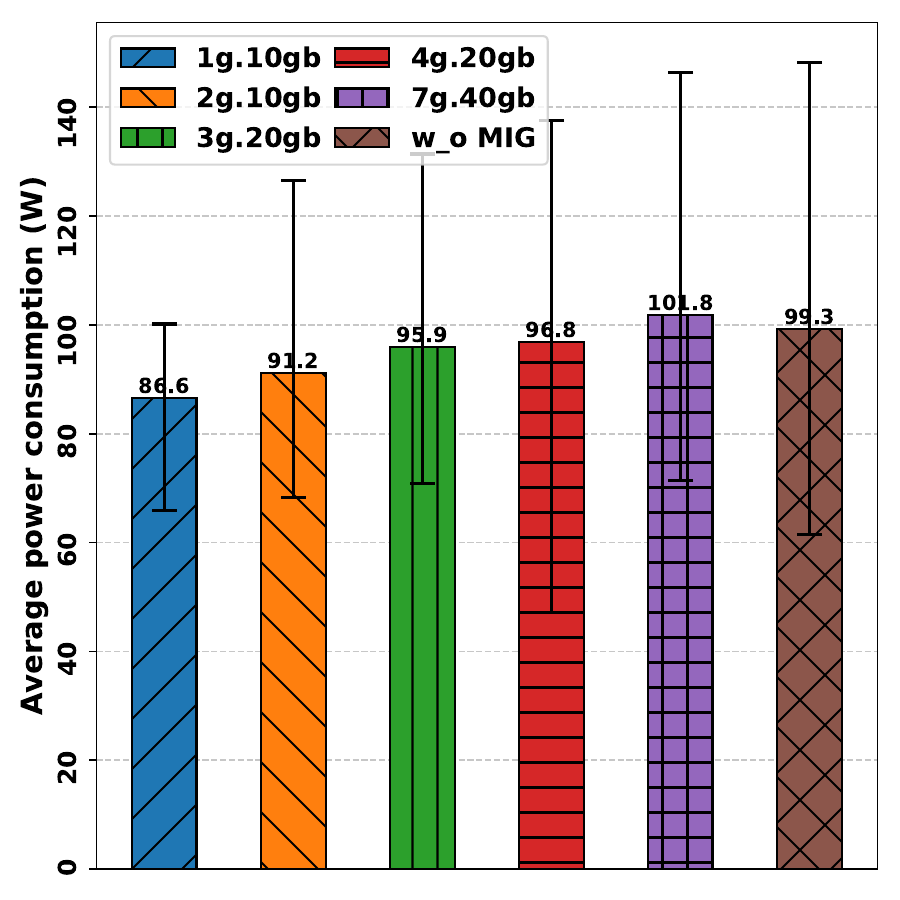}
        \caption{Avg. power consumption (W)}
        \label{fig:MIG_C}
    \end{subfigure}
    \caption{Figure of the results showing the effect of different memory sizes and compute units on the NVIDIA A100 using MIG for the inference of the ConvNeXt-Large network with a batch size of 1, illustrating the impact in terms of throughput, memory usage, and average power consumption.}
    \label{fig:MIG_tab}
\end{figure*}

The results of the benchmark performed on MIG, as shown in Figure \ref{fig:MIG_tab}, allow us to observe that, on the one hand, throughput (Figure \ref{fig:MIG_A}) increases linearly with the number of GPCs allocated. This behavior is expected, as the ConvNeXt Large network is a neural network model that can utilize all available GPU resources; therefore, as more resources are assigned, its performance improves. In other cases, as demonstrated in previous studies \cite{tam}, smaller models may underutilize the available GPU resources; thus, an increase in computational resources does not necessarily lead to a linear improvement in performance. From this figure, it can also be observed that, although the effect is small, using MIG does have an impact on performance during model inference: the throughput is slightly lower when using a MIG partition equivalent to the full GPU compared to using the GPU without MIG.

Regarding memory usage, as shown in Figure \ref{fig:MIG_B}, a very different behavior can be detected between using MIG and not using it. The slight increase in memory usage as more resources are assigned to the partitions, considering that the memory related to the network and input is constant in all cases, is due to the use of the internal memory of the SMs, which is accounted for in the total memory used but is restricted by the allocation of the GPCs. What seems clear is that by using MIG, the device is forced to have a more optimized memory allocation than if inferences are executed without it. Despite this, no definitive conclusion can be drawn from this experiment alone, and a deeper investigation lies beyond the scope of this work. Nevertheless, this behavior is consistent with observations reported in similar studies employing this technology \cite{zhang2023migperfcomprehensivebenchmarkdeep}.

Finally, in terms of power consumption (Figure \ref{fig:MIG_C}), it can be seen that MIG presents a potential as a control element for the maximum power consumed, depending on the assigned resources, obviously at the cost of system performance, as has already been demonstrated. It is worth mentioning that, as with throughput, the use of MIG also entails an overhead, as can be seen from the partition equivalent to the GPU having slightly lower consumption compared to using the GPU without this technology.

\subsection{Results of the partitioning performance impact employing GC}
\begin{figure*}[htbp]
    \centering
    \begin{subfigure}[t]{0.30\textwidth}
        \centering
        \includegraphics[width=\textwidth]{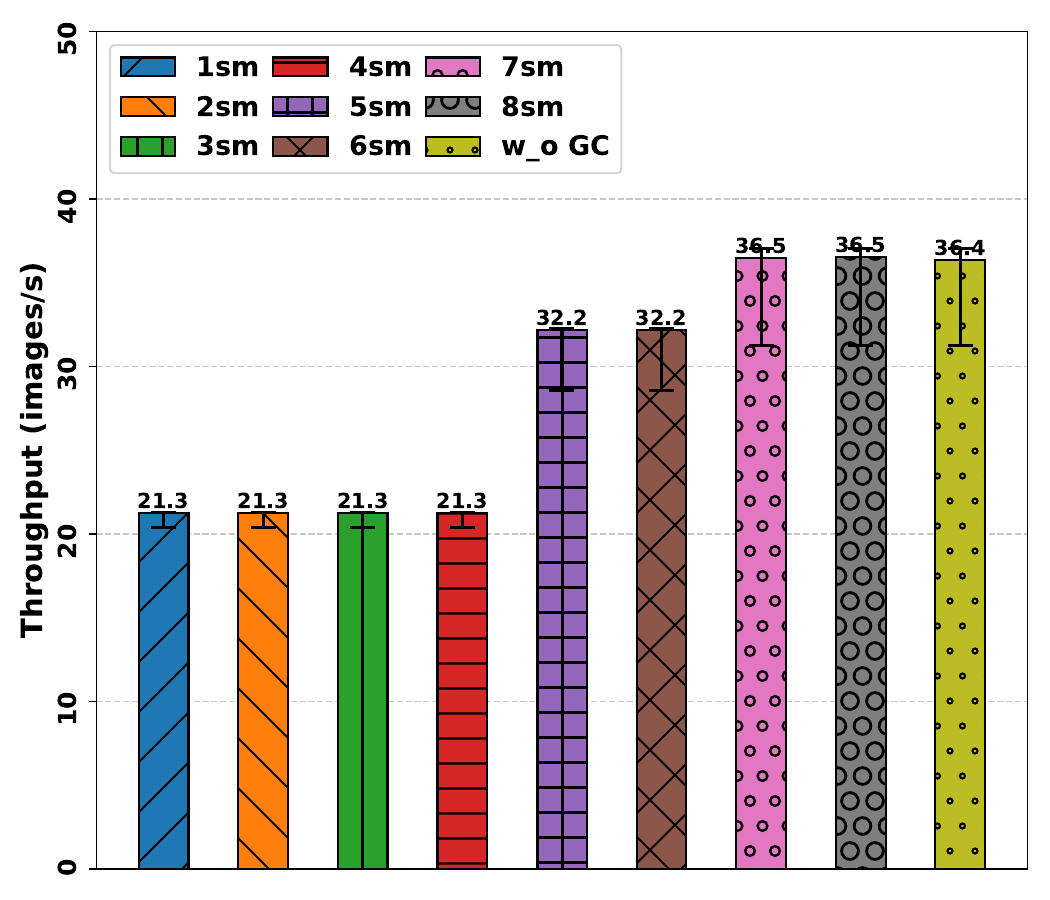}
        \caption{Throughput (images/s)}
        \label{fig:GC_A}
    \end{subfigure}
    \hfill
    \begin{subfigure}[t]{0.30\textwidth}
        \centering
        \includegraphics[width=\textwidth]{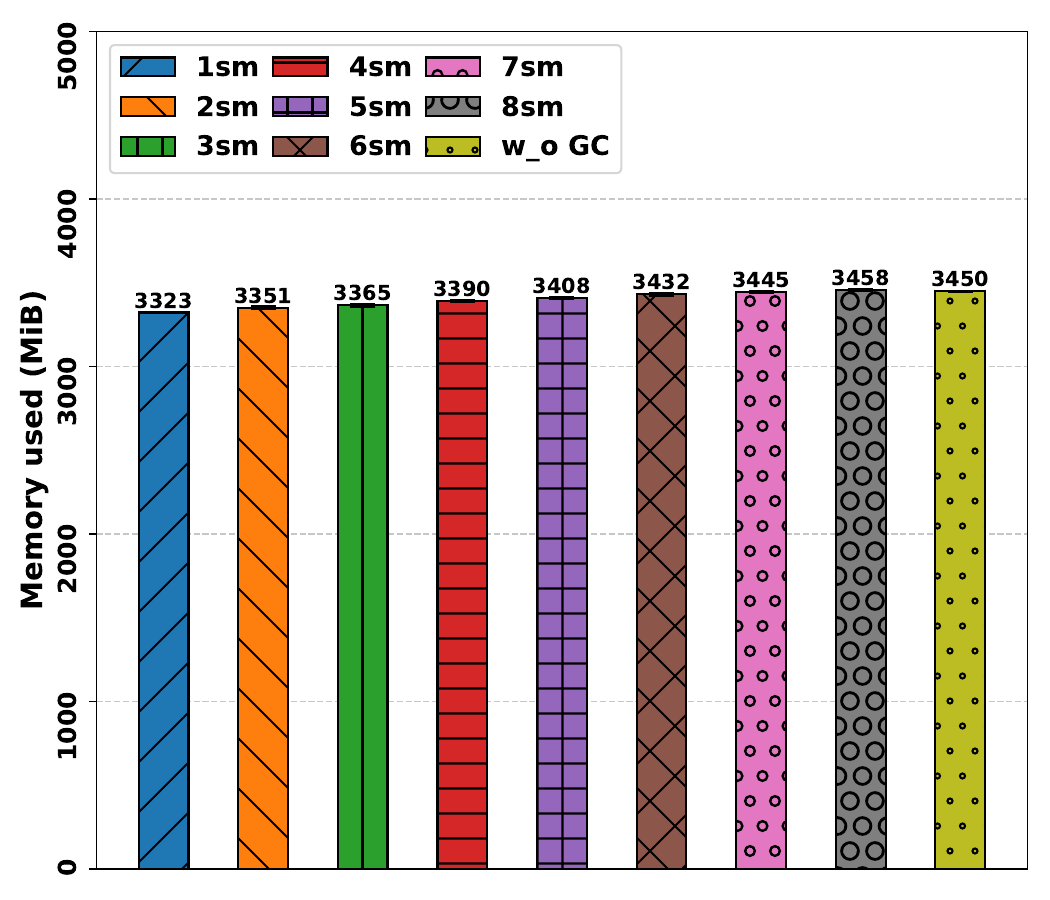}
        \caption{Memory used (MiB)}
        \label{fig:GC_B}
    \end{subfigure}
    \hfill
    \begin{subfigure}[t]{0.30\textwidth}
        \centering
        \includegraphics[width=\textwidth]{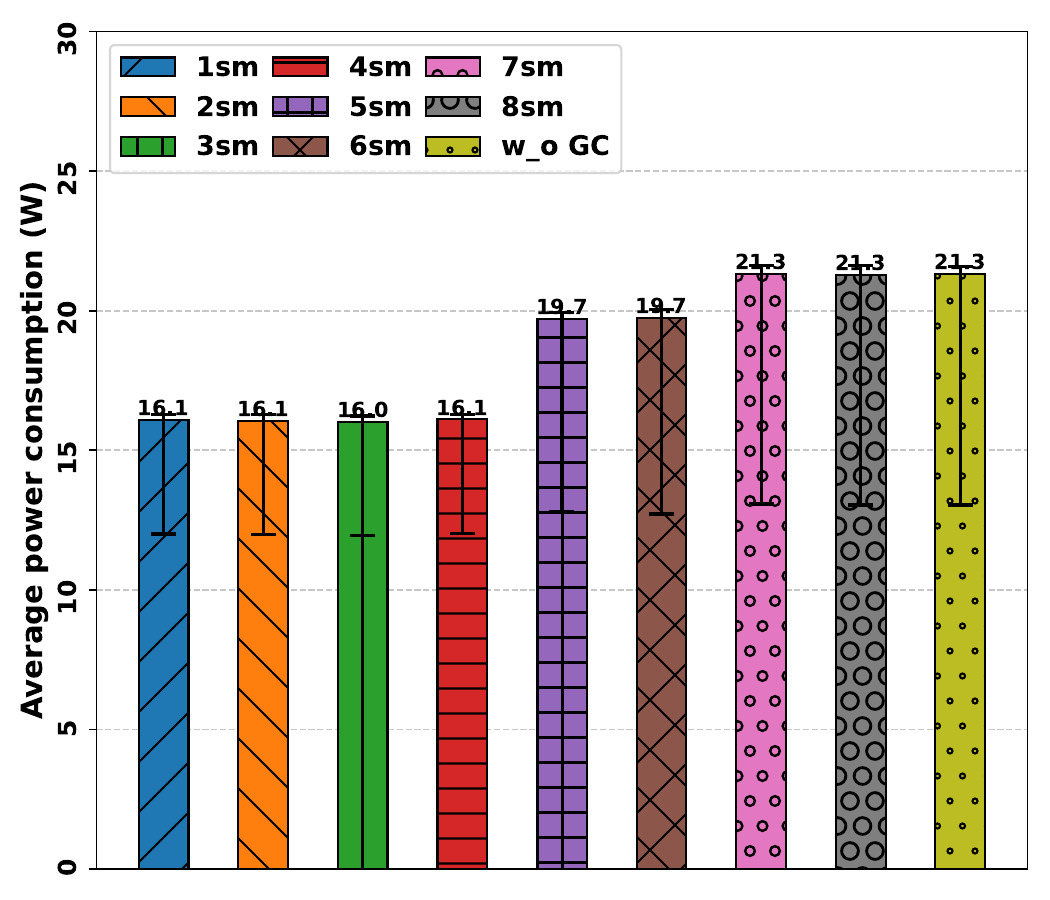}
        \caption{Avg. power consumption (W)}
        \label{fig:GC_C}
    \end{subfigure}
    \caption{Figure of the results showing the effect of different SM assignments on the Jetson Orin Nano using GC for the inference of the ConvNeXt-Large network optimized, with a batch size of 1, illustrating the impact in terms of throughput, memory usage, and average power consumption.}
    \label{fig:GC}
\end{figure*}

Regarding the results obtained from the benchmark using Green Contexts, as shown in Figure~\ref{fig:GC}, a distinct behavior in terms of throughput can be observed, as further illustrated in Figure~\ref{fig:GC_A}. Specifically, although the number of assigned SMs increases linearly, the corresponding performance improvement does not scale proportionally. This behavior confirms that the Jetson devices are based on a compute architecture of version 8.X, as described in the documentation and summarized in Table~\ref{tab:gc_sm_allocation}. The minimum number of assignable SMs is 4, and the allocation can only increase in multiples of 2. Consequently, this explains the throughput behavior observed in the experiments and also indicates that, at least on the Jetson Orin Nano, only two Green Contexts can be executed in parallel, as the device includes only 8 SMs and therefore supports only two non-overlapping partitions.

Additionally, it is worth noting that, unlike MIG, Green Contexts do not appear to introduce any measurable performance overhead, at least in terms of throughput.

As expected, and given that Green Contexts do not provide any form of memory isolation, memory usage remains nearly identical across all configurations, as shown in Figure~\ref{fig:GC_B}. The slight variations observed are likely attributable to the private L0 memory associated with each SM.

Finally, regarding average power consumption, Figure~\ref{fig:GC_C} shows that, similarly to MIG, the number of assigned SMs directly affects the overall power usage of the system. This suggests that Green Contexts could also serve as a mechanism for dynamically limiting power consumption. However, it is worth highlighting that, unlike the results reported for the A100, the power measurements on the Jetson platform are considerably more stable across all configurations, likely due to a higher average GPU utilization.

\subsection{Maximum IMS experimental results}

By applying the algorithm \ref{alg:maxfreq}, described previously, the maximum IMS values were obtained for both the NVIDIA A100 and the Jetson Orin Nano. These results are presented in Tables \ref{tab:fps_max_a100} and \ref{tab:fps_max_orin}. As expected, the IMS values measured on the NVIDIA A100 are consistently higher than those obtained on the Jetson Orin Nano, as the former provides significantly greater computational capacity.

\begin{table}[htbp]
\caption{Maximum IMS values, obtained through an iterative process, on the NVIDIA A100 using the entire GPU stand-alone for the baseline example and MPS, and employing the 3g.20 GB partition for MIG.}
\label{tab:fps_max_a100}
\begin{center}
\begin{tabular}{|c|c|c|c|}
\hline
\multicolumn{4}{|c|}{\textbf{Maximum IMS values for A100 technologies}} \\ 
\hline
 & \textbf{GPU stand-alone}  & \textbf{MPS} & \textbf{MIG (3g 20GB)} \\ 
\hline
ConvNeXt Base & 61 & 61 & 54 \\ 
\hline
ConvNeXt Large & 52 & 52 & 52\\ 
\hline
MobileNetV2 & 147 & 147 & 123 \\ 
\hline
ResNet18 & 129 & 129 & 118 \\ 
\hline
Vit b 16 & 95 & 95 & 76 \\ 
\hline
Vit l 32 & 55 & 55 & 55 \\ 
\hline
\end{tabular}
\end{center}
\end{table}

Regarding the results obtained on the NVIDIA A100, the behavior of the larger models, such as ConvNeXt Large and ViT L 32, is particularly noteworthy. These models maintain their maximum inference frequency even when the amount of allocated resources is reduced \cite{cnn_efecto}. This can be explained by the fact that large models are typically compute-bound, meaning they exhibit a high ratio of arithmetic operations to memory accesses and possess sufficient intrinsic parallelism to continue saturating the streaming multiprocessors (SMs). Consequently, their performance degrades only gradually

In contrast, smaller models tend to be memory-bound: they employ lighter kernels, perform fewer operations per layer, and exhibit lower intrinsic parallelism. As a result, they depend more heavily on GPU scheduling and memory bandwidth. When the number of available SMs or the assigned priority is reduced, these models are unable to hide latency or maintain efficiency, leading to a steeper decline in their maximum inference frequency and a higher likelihood of timeouts.

On the other hand, the behavior described above is slightly altered in the case of the Jetson Orin Nano. Although the percentage reduction in resources is identical (i.e., approximately half of the SMs), the absolute amount of available computational resources is substantially smaller. This results in a reduction in the number of SMs, leading to a more linear decrease in maximum IMS values across all models. In this scenario, the largest models become the most affected in relative terms, showing the highest percentage drop in performance.

\begin{table}[htbp]
\caption{Maximum IMS values, obtained through an iterative process, on the Jetson Orin Nano, using the entire GPU stand-alone for the baseline example and MPS, and employing the 4 SMs partition for GC.}
\label{tab:fps_max_orin}
\begin{center}
\begin{tabular}{|c|c|c|c|}
\hline
\multicolumn{4}{|c|}{\textbf{Maximum IMS values for Jetson technologies}} \\ 
\hline
 & \textbf{GPU stand-alone}  & \textbf{MPS} & \textbf{GC (4 SMs)} \\ 
\hline
ConvNeXt Base & 58 & 58 &  35\\ 
\hline
ConvNeXt Large & 33 & 33 & 18 \\ 
\hline
MobileNetV2 & 134 & 134 &  79\\ 
\hline
ResNet18 & 127 & 127 &  70\\ 
\hline
Vit b 16 & 58  & 58 & 36\\ 
\hline
Vit l 32 & 42 & 42 & 24\\ 
\hline
\end{tabular}
\end{center}
\end{table}

\subsection{Experimental results of two processes inferring in parallel}
Given the results obtained, a very similar isolation behavior can be observed as a function of network size, with consistent performance patterns appearing across pairs of models. MobileNetV2 and ResNet18 represent the group of smaller, less complex models; ConvNeXt Base and ViT B 16 correspond to medium-sized, moderately complex models; and ConvNeXt Large and ViT L 32 represent the largest and most computationally demanding models.

Therefore, to improve the clarity of this section and avoid redundant information that does not provide additional insights, only one representative model from each group is presented:

\subsubsection{Isolation results between two processes on the A100}

\begin{figure*}[h]
    \centering
    \begin{subfigure}[b]{0.33\textwidth}
        \includegraphics[width=\textwidth]{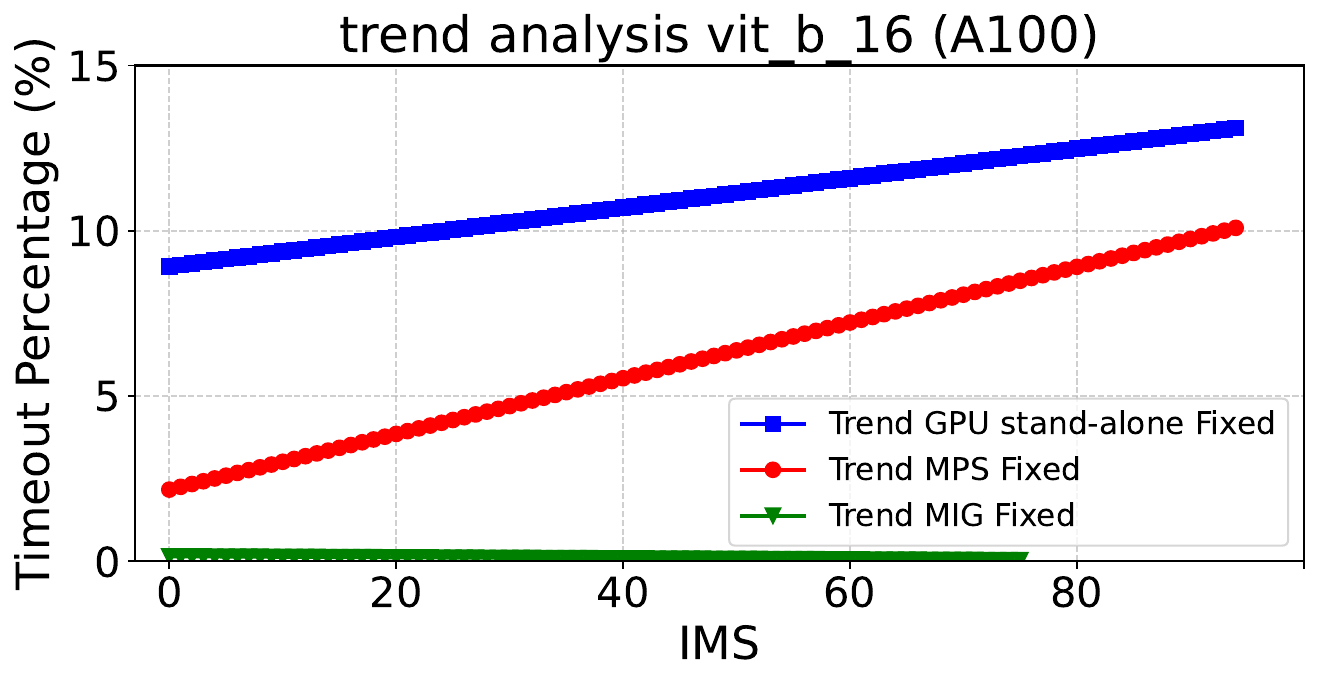}
        \caption{Vit\_b\_16 (A100)}
        \label{fig:A100_vitb16}
    \end{subfigure}%
    \begin{subfigure}[b]{0.33\textwidth}
        \includegraphics[width=\textwidth]{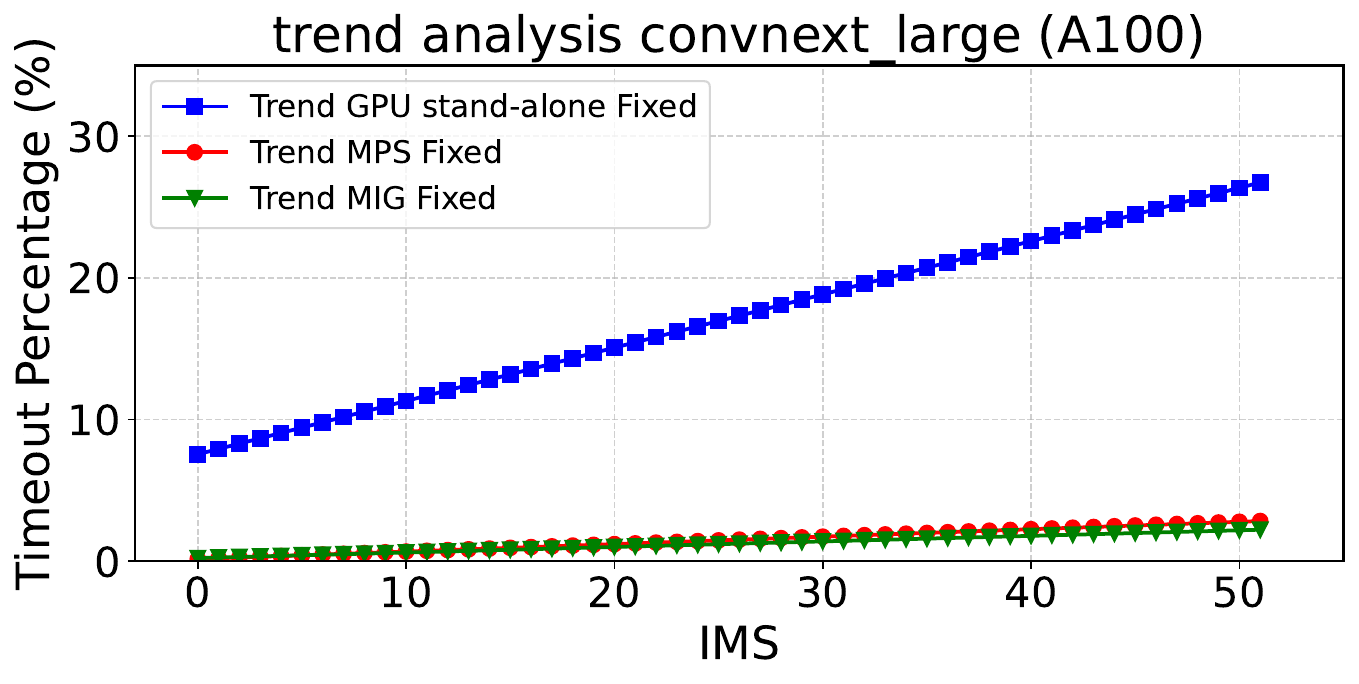}
        \caption{ConvNeXt-Large (A100)}
        \label{fig:A100_convnext}
    \end{subfigure}%
    \begin{subfigure}[b]{0.33\textwidth}
        \includegraphics[width=\textwidth]{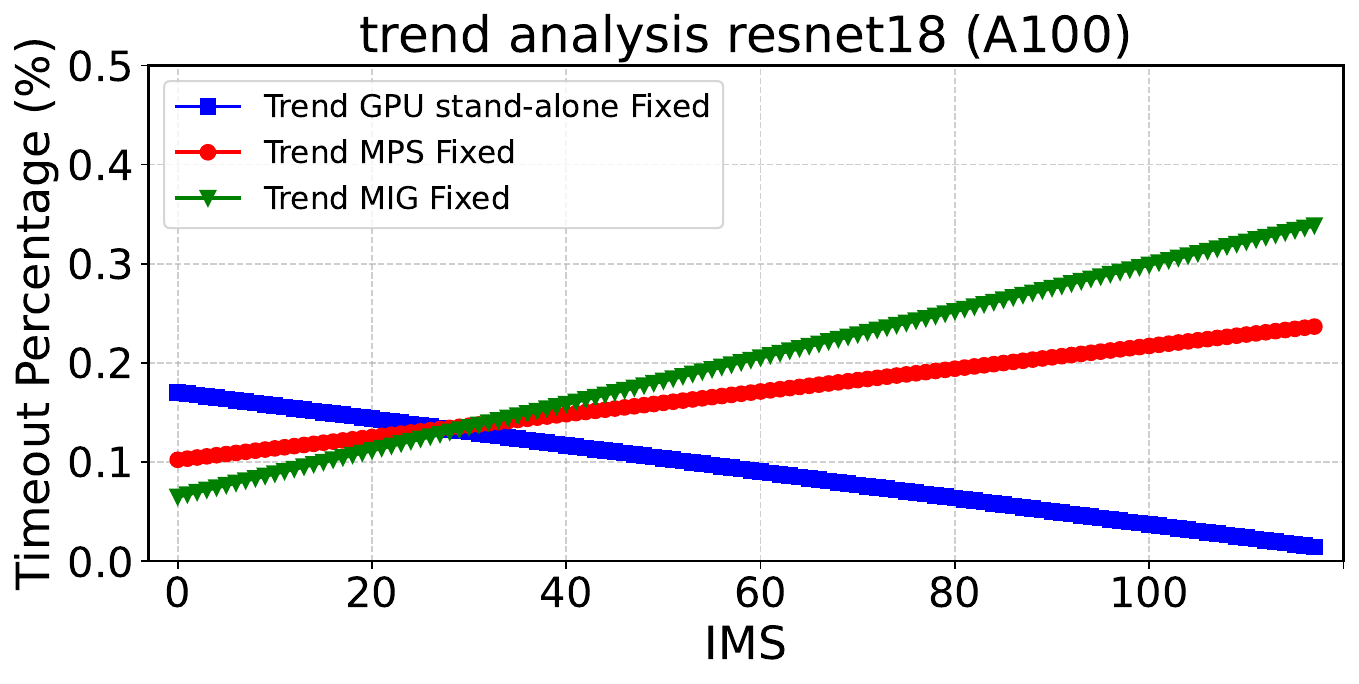}
        \caption{ResNet18 (A100)}
        \label{fig:A100_resnet}
    \end{subfigure}
    \caption{Percentage of inferences detected as \textit{timeouts} relative to the total number of inferences, in the case of two parallel processes running the same network. Only the trend of the process performing inference at its calculated maximum frequency (Fixed) is shown, using GPU stand-alone, MPS, and MIG on the Nvidia A100.}
    \label{fig:A100_unificado}
\end{figure*}

The results are presented using graphs that illustrate the trend in timeout occurrences, with each figure showing the behavior of the three evaluated technologies for the same model. After analyzing the results obtained from the experiments shown in Figure \ref{fig:A100_unificado}, the observed behavior is generally as expected.

For the medium-sized model (Figure \ref{fig:A100_vitb16}), the observed behavior aligns with expectations. The stand-alone GPU configuration exhibits the lowest performance, which is significantly improved when MPS is employed. Regarding MIG, the results are highly favourable, as the number of timeouts remains close to zero across all IMS values.

On the other hand, for the larger models (Figure \ref{fig:A100_convnext}), the most notable observation is the substantial performance improvement achieved with MPS when compared to the results obtained with MIG. This behavior is consistent with previous observations, as the primary purpose of MPS is not to provide strict resource isolation but rather to create unified contexts that reduce the cost of context switching between processes. Consequently, larger models, which are most affected by context switching, also show the most significant performance gains under MPS. Furthermore, in this case study, the first timeouts appear despite the use of MIG at higher frequencies, highlighting that MIG does not provide complete isolation.

Finally, for the smallest models (Figure \ref{fig:A100_resnet}), the results illustrate a previously noted phenomenon: the overhead associated with MIG relative to the other alternatives, which makes it the least efficient option in terms of isolation for this model. Nevertheless, the magnitude of timeouts for this model is minimal and is likely due to inherent variability in inference times when operating at very high frequencies. Therefore, it can be concluded that all three alternatives are suitable for inferring such small models, and in all cases, the available GPU resources are likely underutilized.

\subsubsection{Isolation results between two processes on jetson orin Nano}

\begin{figure*}[h]
    \centering
    \subfloat[Vit\_b\_16 (Jetson orin Nano)]{\includegraphics[width=0.33\textwidth]{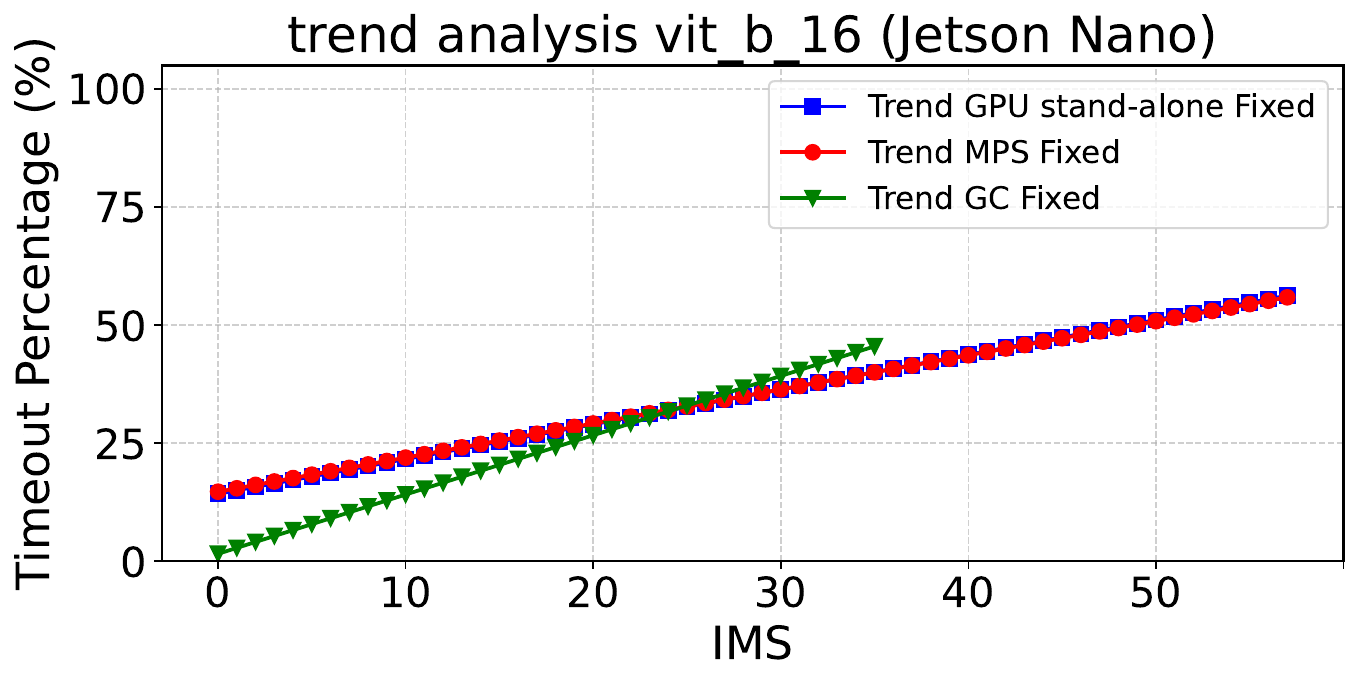}}
    \subfloat[ConvNeXt-Large (Jetson orin Nano)]{\includegraphics[width=0.33\textwidth]{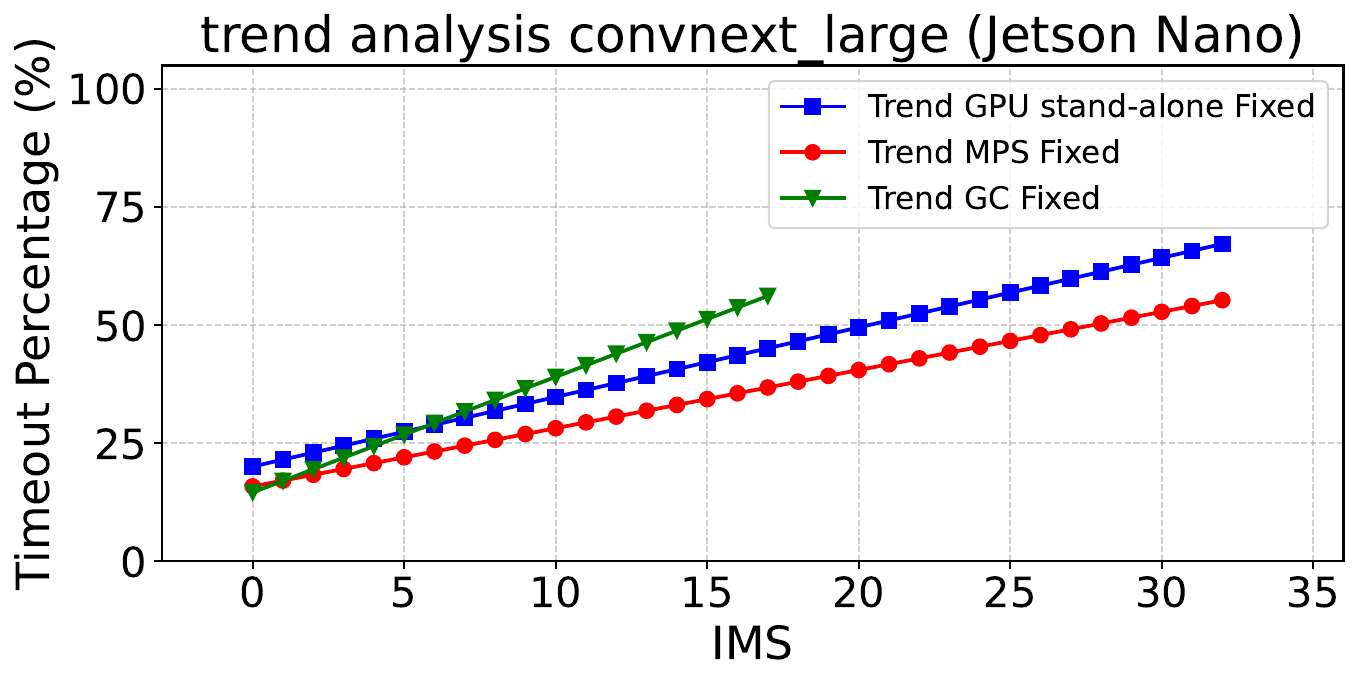}}
    \subfloat[Resnet18 (Jetson orin Nano)]{\includegraphics[width=0.33\textwidth]{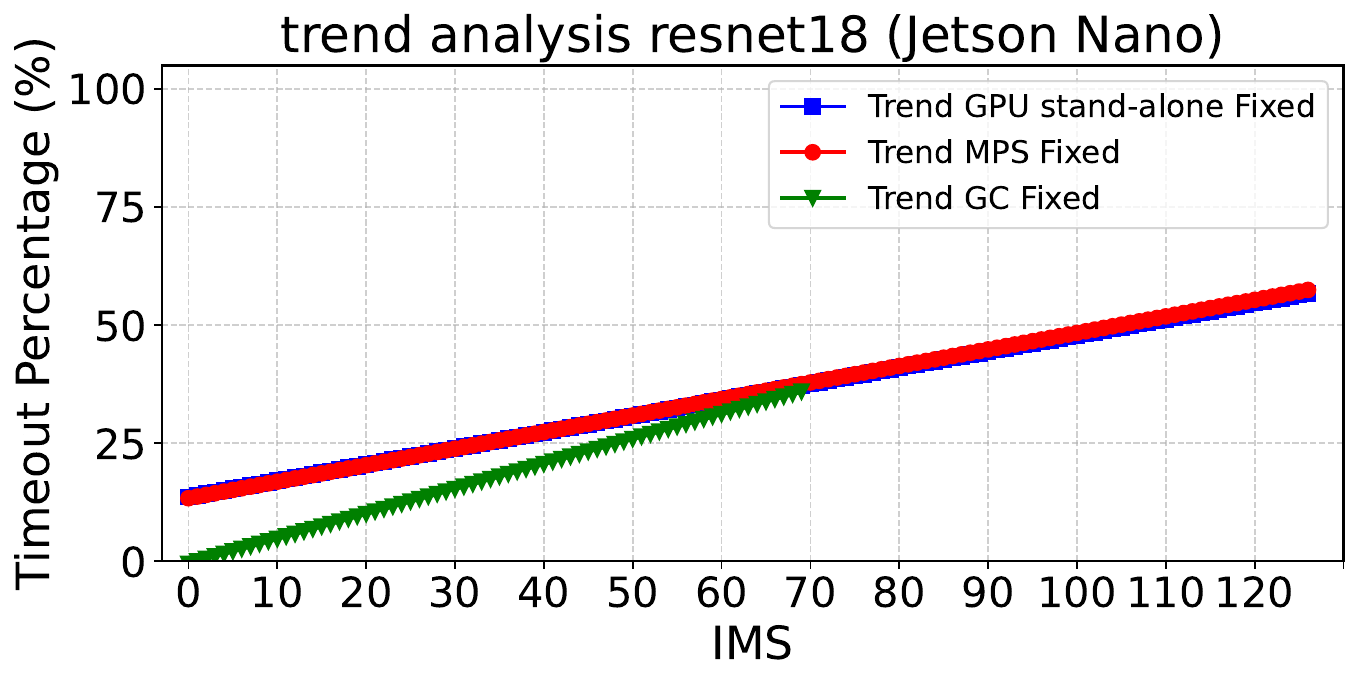}}
    \caption{Percentage of inferences detected as \textit{timeouts} relative to the total number of inferences, in the case of two parallel processes running the same network. Only the trend of the process performing inference at its calculated maximum frequency (Fixed) is shown, using GPU stand-alone, MPS, and GC on the  Jetson orin Nano.}
    \label{fig:nano_unificado}
\end{figure*}

After repeating the previous experiments on the Jetson Orin Nano, it can be observed in Figure \ref{fig:nano_unificado} that the three networks exhibit very similar behavior in terms of timeouts, regardless of their size, which would not be an expected behavior given the size differences between them. Regarding the comparison between GPU and MPS, both technologies exhibit very similar behavior, except that, as previously observed, using the larger models yields some improvement, albeit much smaller than what was seen with the A100.

\begin{figure}[h]
\centering
\includegraphics[width=0.43\textwidth]{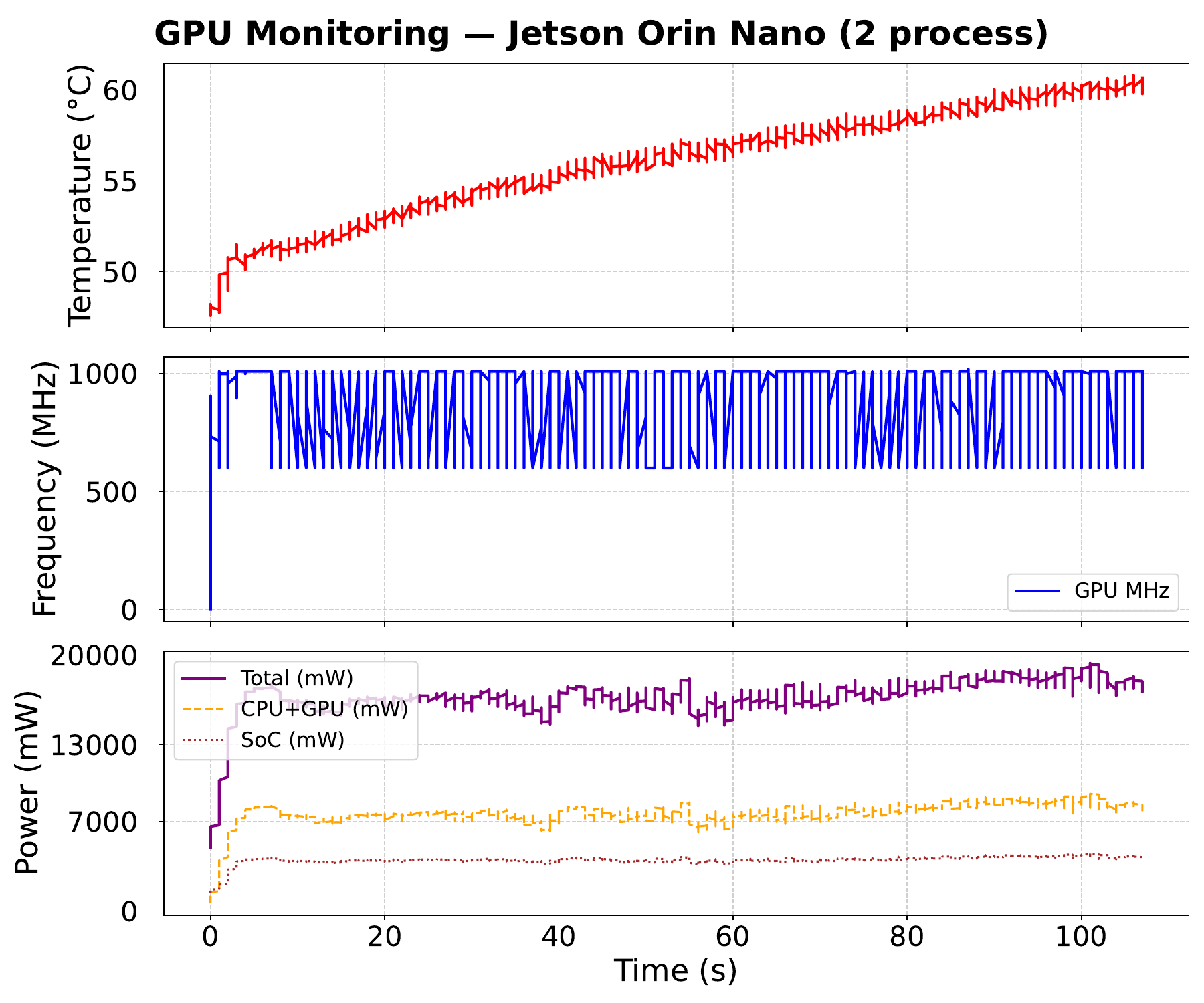}
\caption{Graph analyzing the evolution of the Jetson Orin Nano GPU under two parallel inference processes in terms of temperature, frequency, and power consumption.}
\label{fig:2pro}
\end{figure}

Additionally, the results obtained with GC, which aims to emulate MIG-like behavior in the Jetson environment, do not appear to provide the expected level of isolation typically offered by strict resource partitioning technology. This necessitates investigating whether this behavior is due to a malfunction of the technology or some other hardware limitation. As shown in Figure \ref{fig:2pro}, which illustrates the GPU frequency, power consumption, and temperature of the Jetson Orin Nano during the execution of selected experiments, the GPU frequency undergoes sudden drops to values close to half of its maximum. After inspecting the power consumption, this behavior appears to be caused by operating two models in parallel, utilizing the complete set of available GPU resources. Under these conditions, the device approaches a power draw of around 20 W, the maximum supported by this model in its highest-power mode, suggesting that the system is entering a saturation state. This likely explains the drastic frequency reductions and the corresponding performance degradation observed in the experiments.

Therefore, these results indicate that GC cannot be considered a conclusive isolation technique for this device, given the inherent hardware limitations of the Jetson Orin Nano.

It is also worth highlighting an interesting behavior observed when comparing Jetson-class platforms with general-purpose GPUs: in terms of inference time, Jetson platforms exhibit noticeably greater stability. This is because Jetson devices are built on an integrated architecture in which the CPU, GPU, and memory share a tightly optimized environment designed explicitly for inference workloads. Such integration reduces variability caused by dynamic resource management and internal contention, factors that can occur in general-purpose GPUs, which rely on less predictable operating systems and driver layers \cite{efecto_inf}. This effect is evident when observing that the variation in timeouts, as the frequency increases, does not show the drastic fluctuations seen in the A100 results.

\subsubsection{Isolation Results Between Two Processes on Jetson Orin AGX (Equivalent to Orin Nano)}

Based on the results related to GC discussed in the previous section, we explored the possibility of repeating the same experiments in a scenario where power consumption is not a limiting factor. Therefore, the first step was to replace the experimental platform from a Jetson Orin Nano to a Jetson Orin AGX. The Jetson AGX features the same chip architecture but offers double the maximum power, reaching 50 W in its peak mode. It also provides a GPU with a higher maximum frequency of 1.3 GHz, increased memory bandwidth, and a total of 16 SMs.

Nevertheless, in order to ensure that the platform change does not drastically affect the results, the per-process SM allocation was maintained at 4 SMs, and the GPU frequency was fixed at 1.02 GHz, which corresponds to the maximum frequency achievable by the Orin Nano.

With the characteristics described previously, in this case, unlike the previous ones, since no comparison is being made between the three technologies, the number of timeouts of the process performing inference at a fixed frequency will be shown. This corresponds to the process of interest that has been graphed in the previous figures, compared to the process whose frequency is being adjusted. Additionally, in this case, the maximum inference frequency for the adjustable process has even been increased above the value that ensures compliance with the inference time, in order to guarantee isolation despite detecting timeouts in the background process.

\begin{figure*}[h]
    \centering
    \subfloat[Vit\_b\_16 (Jetson orin AGX)]{\includegraphics[width=0.33\textwidth]{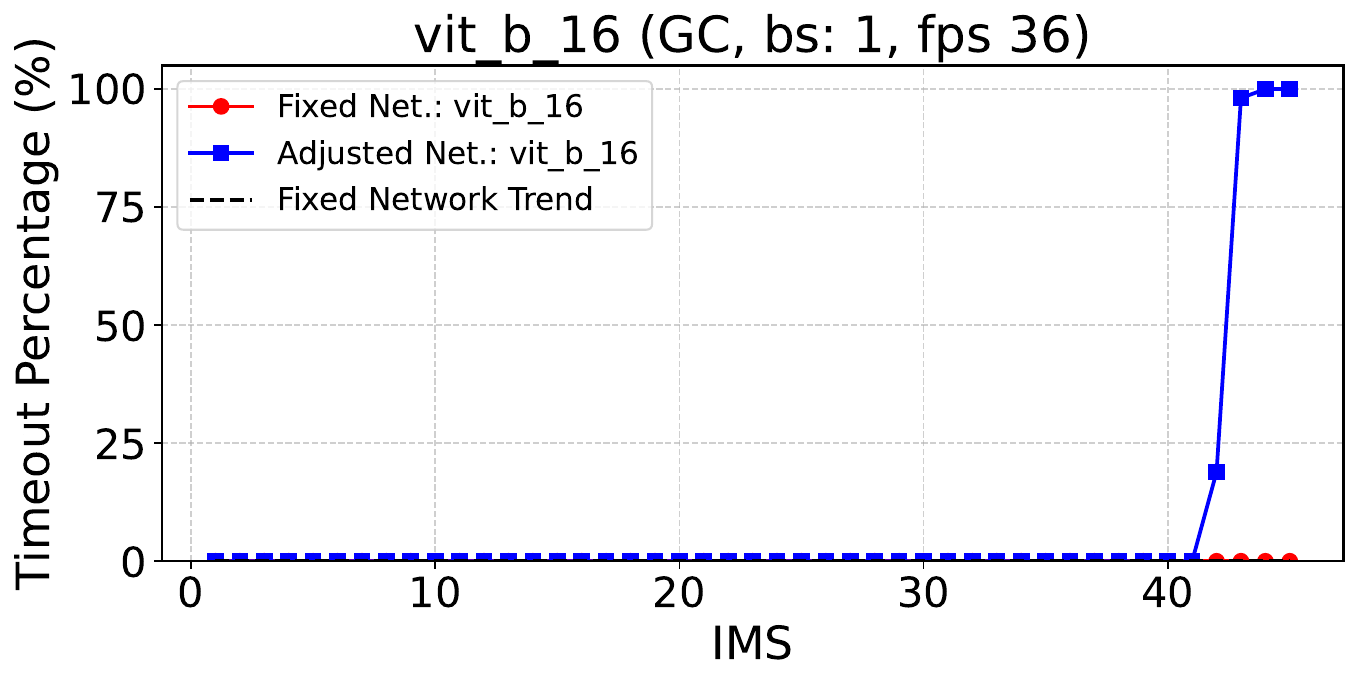}}
    \subfloat[ConvNeXt-Large (Jetson orin AGX)]{\includegraphics[width=0.33\textwidth]{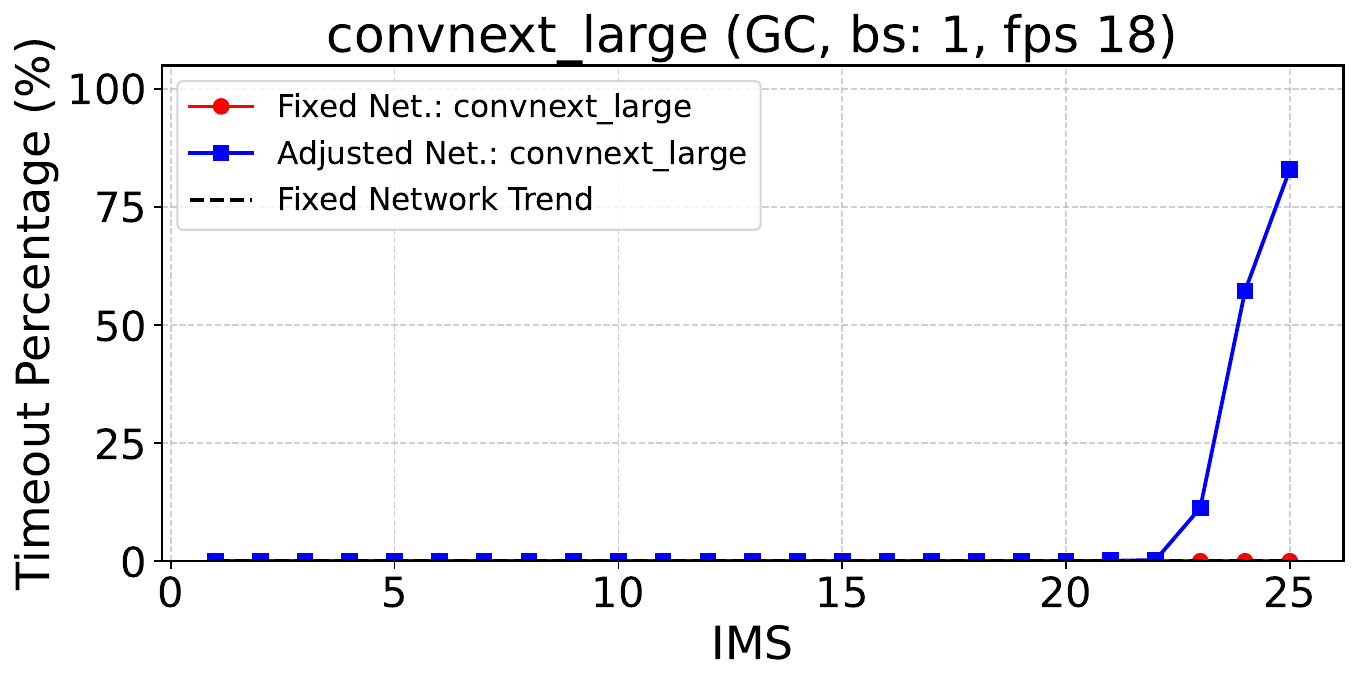}}
    \subfloat[Resnet18 (Jetson orin AGX)]{\includegraphics[width=0.33\textwidth]{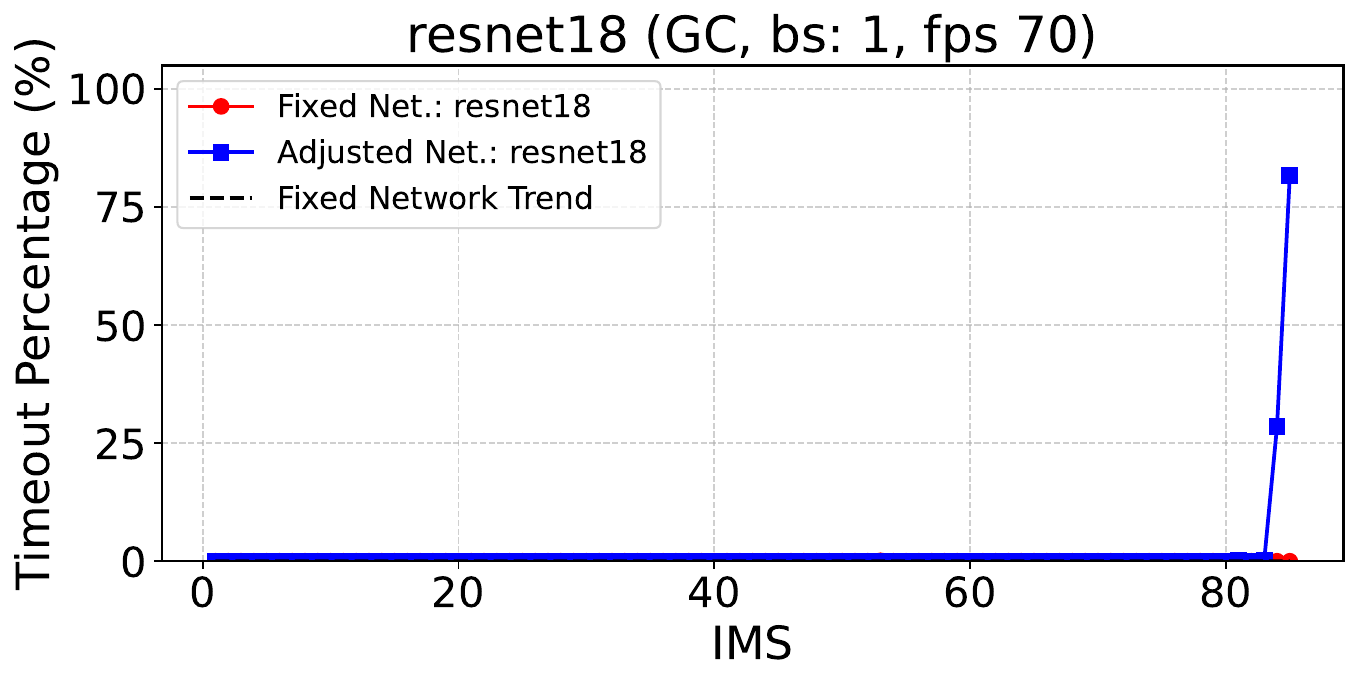}}
    \caption{Percentage of inferences detected as \textit{timeouts} relative to the total number of inferences, in the case of two parallel processes running the same network: one performing inference at its calculated maximum frequency (Fixed) and the other progressively increasing its inference rate (Adjusted), using GC on the Jetson Orin AGX, fixing the frequency at 1.02 GHz.}
    \label{fig:AGX}
\end{figure*}

As can be observed in the graphs presented in Figure \ref{fig:AGX}, the percentage of timeouts is drastically reduced in all cases. This improvement occurs because, unlike the previous scenario, power consumption limitations no longer restrict the GPU’s maximum frequency, achieving nearly ideal process isolation, very similar to the isolation provided by MIG on the A100.

\subsubsection{Results on Jetson Orin AGX for Four Processes}

Finally, as the last comparative study, an experiment is conducted that, similarly to the tests performed on the Nano board, utilizes all the resources available on the Orin AGX. The problem is that, unlike GC, the remaining approaches do not allow reproducing the behavior of the Jetson Orin Nano on the AGX, since it is not possible to restrict the amount of resources they use. Therefore, in order to fairly compare these three technologies on this platform, we opted to execute four processes in parallel, given that the AGX has twice the compute resources. In the case of GC, a configuration of four partitions, each with four SMs, is used.

Thus, the experiment follows the same structure as before. However, this time it involves four processes: three of them continuously performing inference at their maximum IMS, while the fourth process, similarly to the previous case, progressively increases its inference request rate. This setup allows evaluating the level of isolation across the four partitions, which remain configured as four separate SMs.

\begin{figure}[h]
\centering
\includegraphics[width=0.43\textwidth]{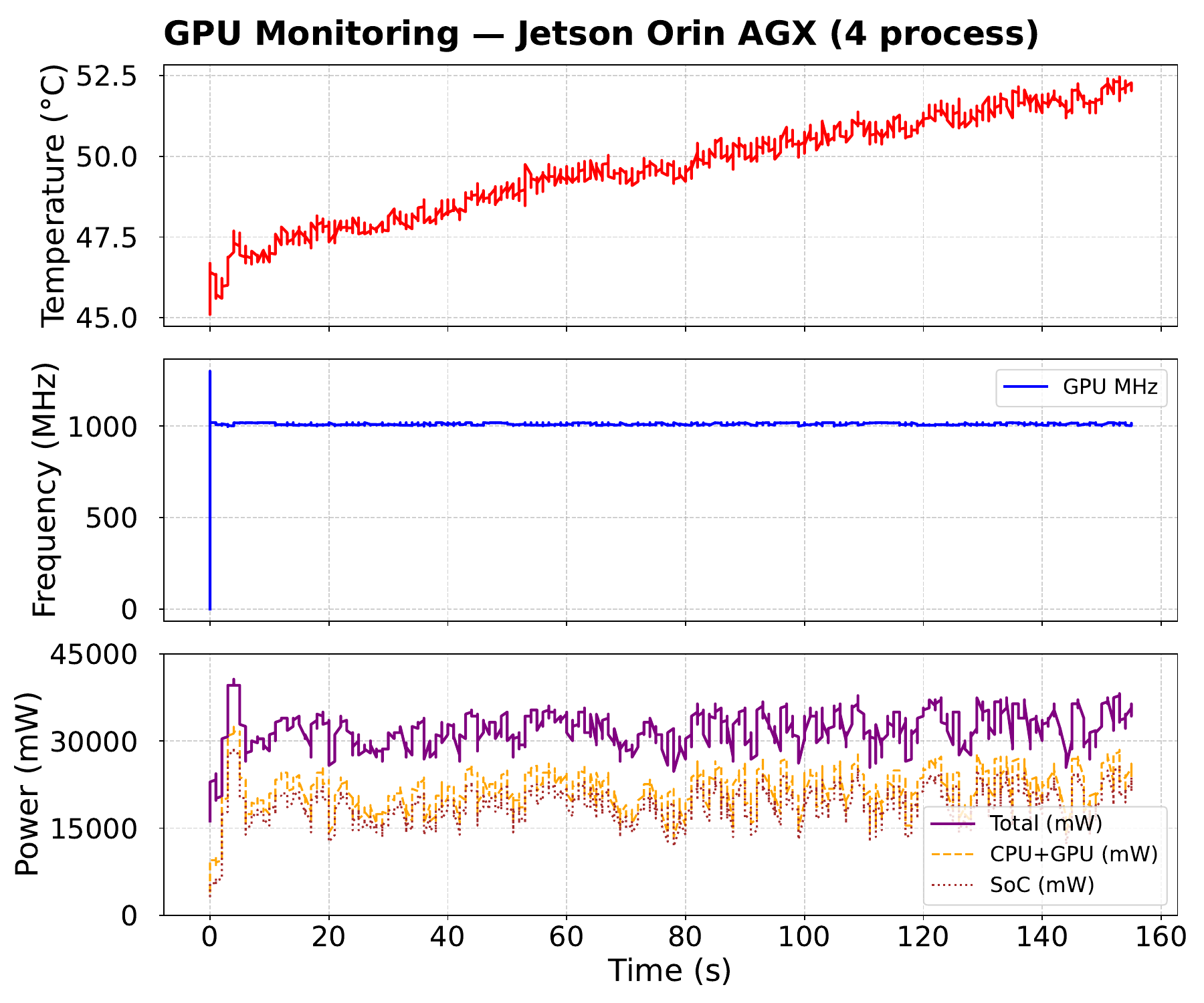}
\caption{Graph analyzing the evolution of the Jetson Orin AGX GPU under four parallel inference processes in terms of temperature, frequency, and power consumption.}
\label{fig:4pro}
\end{figure}

Before conducting the isolation benchmarks, the analysis of temperature, frequency, and power consumption is replicated for the four-process scenario to verify that the hardware limitations previously assessed are not being exceeded. As shown in Figure \ref{fig:4pro}, the GPU frequency remains stable, the temperature increase during execution is more moderate, and power consumption never approaches the system’s theoretical maximum. This explains the significant theoretical differences previously observed between MPS and GC compared to the results obtained on the A100.

\begin{figure*}[h]
    \centering
    \begin{subfigure}[b]{0.33\textwidth}
        \includegraphics[width=\textwidth]{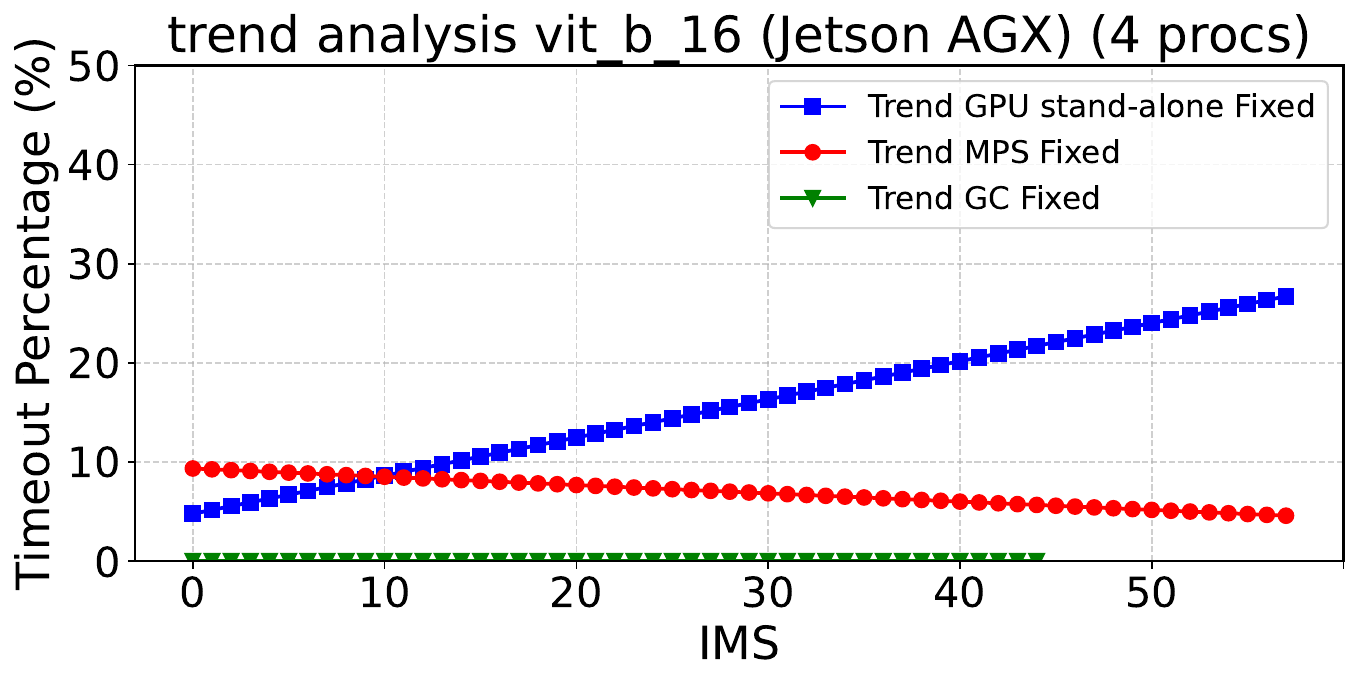}
        \caption{Vit\_b\_16 (Jetson Orin AGX)}
        \label{fig:AGX_vitb16}
    \end{subfigure}%
    \begin{subfigure}[b]{0.33\textwidth}
        \includegraphics[width=\textwidth]{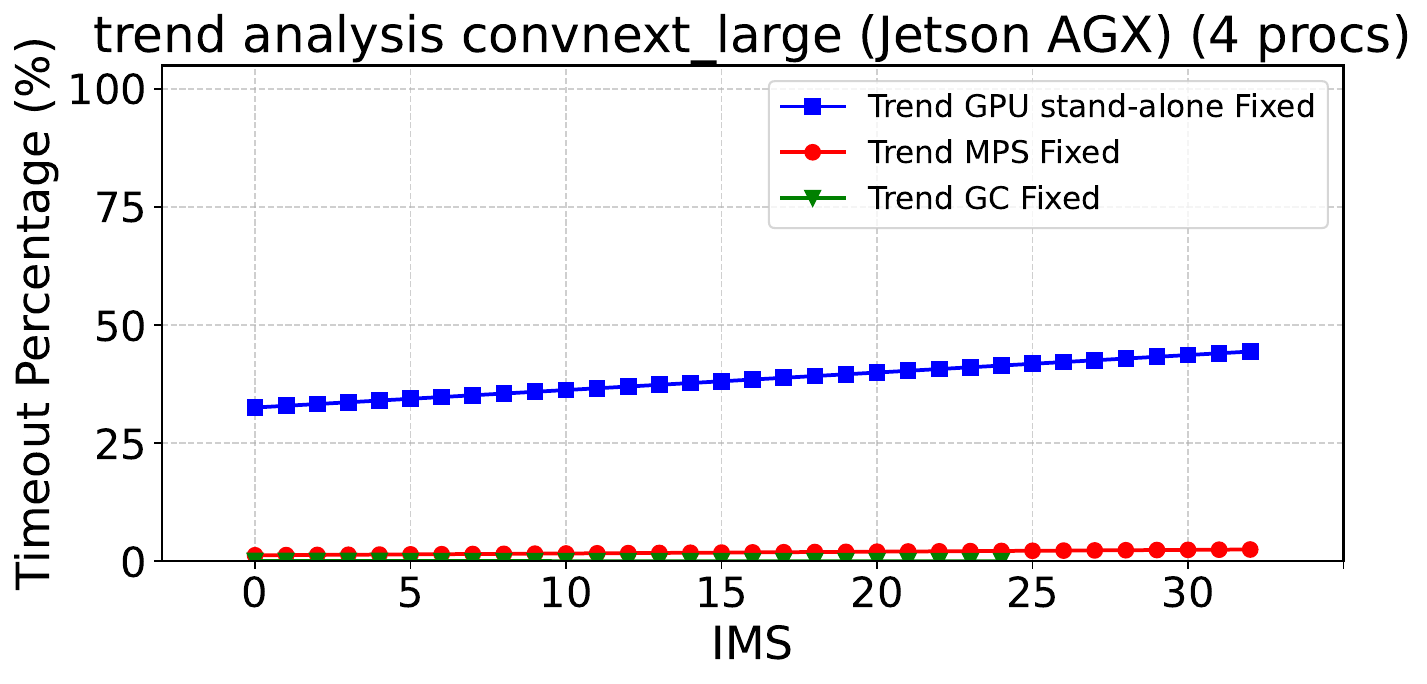}
        \caption{ConvNeXt-Large (Jetson Orin AGX)}
        \label{fig:AGX_convnext}
    \end{subfigure}%
    \begin{subfigure}[b]{0.33\textwidth}
        \includegraphics[width=\textwidth]{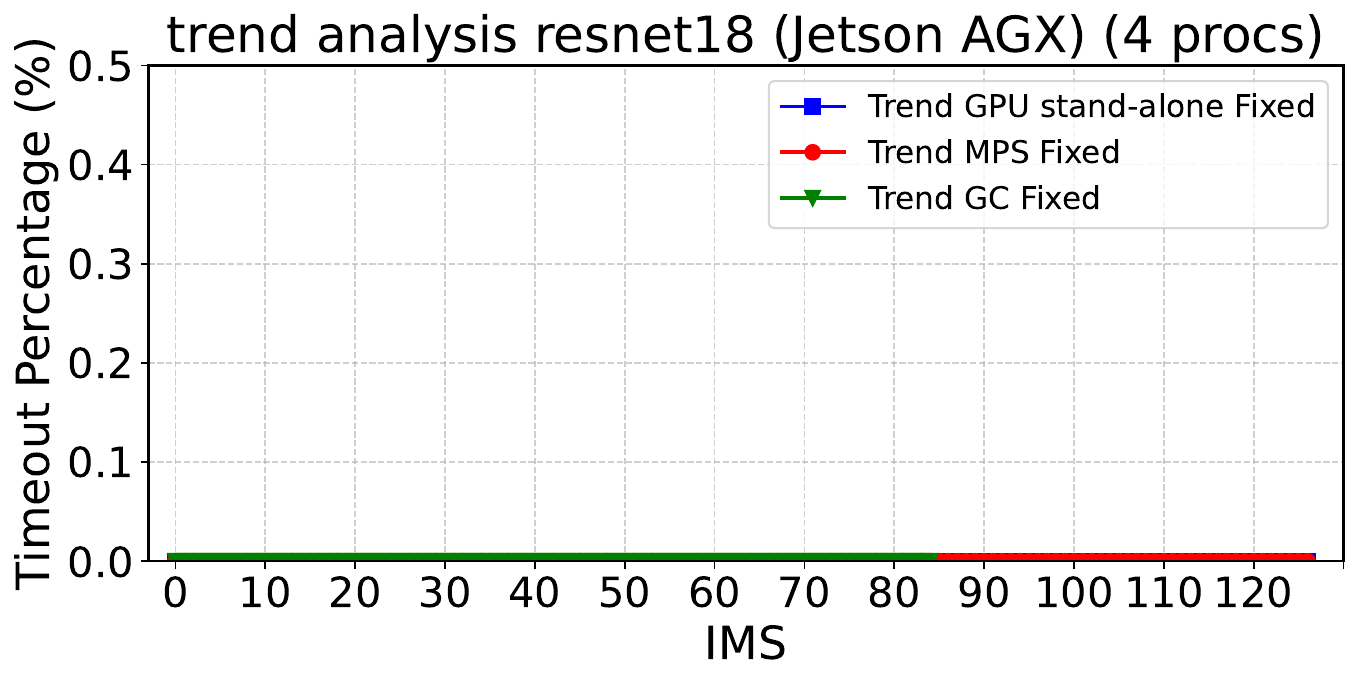}
        \caption{ResNet18 (Jetson Orin AGX)}
        \label{fig:AGX_resnet}
    \end{subfigure}

    \caption{Percentage of inferences detected as \textit{timeouts} relative to the total number of inferences, in the case of four parallel processes running the same network. The plot shows the mean trend of the three fixed processes that perform inference at their calculated maximum (fixed) frequency, using an independent GPU, MPS, and GC on the Jetson Orin AGX.}
    \label{fig:AGX_unificado}
\end{figure*}

After verifying that the platform does not exhibit any hardware-related limitations, the same experiments previously described are carried out, as shown in Figure \ref{fig:AGX_unificado}.

Regarding the results, the behavior of ConvNeXt-Large, shown in Figure \ref{fig:AGX_convnext}, and that of ResNet18, illustrated in Figure \ref{fig:AGX_resnet}, is identical to what was observed in the A100 experiments. For the larger model, both GC and MPS provide improvements very similar to those obtained on the A100, clearly outperforming the stand-alone configuration. On the other hand, in the case of ResNet18, the model does not exhibit interference between processes, as its small size leads to underutilization of the available resources, and therefore, no significant contention occurs. As a result, all three approaches are equally suitable when deploying this network.

The most unusual behavior appears in the case of ViT B 16, presented in Figure \ref{fig:AGX_vitb16}. On one hand, the behavior of the stand-alone GPU and Green Context is as expected and closely resembles that of MIG on the A100. However, in the case of MPS, a slightly atypical trend can be observed, as performance decreases marginally as the frequency is reduced. Nevertheless, since this behavior is isolated and the deviation from the theoretically expected trend is slight, it can be attributed to natural variability in inference time. Therefore, it does not invalidate the feasibility of MPS as a straightforward alternative to the stand-alone GPU configuration.

\section{conclusions}

This work presents a comprehensive analysis of the main isolation mechanisms available in modern GPUs. MPS, MIG, and Green Contexts, to evaluate their suitability for guaranteeing predictable inference times in safety-critical applications. The experimental results obtained on both general-purpose platforms (NVIDIA A100) and edge devices (Jetson Orin Nano and AGX) allow us to derive several robust conclusions.

First, MIG proves to be the most robust mechanism in terms of temporal isolation, providing partitions with strong guarantees at both the computational and memory levels. However, MIG also exhibits an important limitation: its configuration is rigid and does not allow dynamic reallocation or resizing of resources once partitions are created. This lack of flexibility makes it difficult to adapt MIG to workloads with variable real-time requirements, restricting its applicability in scenarios where dynamic behavior is essential. Despite these limitations, MIG remains the mechanism capable of delivering the strongest and most reliable temporal guarantees.

Second, MPS offers moderate improvements over native GPU usage, mainly by reducing context-switching overhead and enabling a more efficient use of hardware resources. While MPS can slightly improve throughput and mitigate some contention effects, it does not provide true isolation, particularly in terms of memory bandwidth or SM scheduling guarantees. Therefore, MPS is not suitable for high-criticality systems; however, it remains a practical and appealing alternative for general-purpose workloads, particularly due to its ease of deployment and ability to operate dynamically without requiring predefined or static resource partitions.

Third, Green Contexts emerge as a promising alternative for edge devices, offering fine-grained control over SM allocation with negligible computational overhead and without noticeable throughput penalties. However, the absence of memory isolation limits their ability to provide complete temporal guarantees. This limitation is particularly evident on power-constrained platforms such as the Jetson Orin Nano, where interactions between power consumption, clock frequency, and contention significantly reduce isolation capabilities. Experiments on the Jetson Orin AGX, however, show that when power is not a limiting factor, Green Contexts can deliver levels of isolation approaching those of MIG, even with multiple concurrent processes.

Ultimately, the study confirms that the nature of the deep learning model has a significant influence on its temporal stability. Compute-bound models, such as ConvNeXt or ViT, show greater resilience to resource reduction, whereas memory-bound models, like ResNet18 or MobileNetV2, exhibit higher sensitivity to contention, especially in environments lacking memory isolation.

Overall, the results demonstrate that current GPU isolation technologies still present limitations in guaranteeing complete temporal predictability across all scenarios; however, they also highlight the potential of more flexible approaches, such as Green Contexts, for embedded systems. The combination of fine-grained partitioning, dynamic resource management, and future developments in memory isolation points toward a next-generation GPU better suited for functional safety systems.

\section{Future work}

This work has served as a starting point to demonstrate the viability of Green Contexts (GC) as a practical alternative to MIG for edge computing systems, and it opens the possibility of applying this technique to general-purpose devices due to its fine-grained resource allocation capabilities.

For future work, there are two clear development directions. On one hand, there is a need to develop memory isolation technology at the context level, enabling GC to operate in parallel while ensuring the same functionalities currently provided by MIG. This would likely improve overall performance as input sizes increase and the number of concurrent processes grows.

On the other hand, similarly to what has already been achieved with technologies like MIG, it is of interest to develop algorithms that allow dynamic resource allocation. Such algorithms would enable the system to meet temporal requirements while maintaining strict isolation between processes dynamically, and could also be employed to control the energy consumption of GPU workloads.

Overall, GC represents a new and promising technology that could emerge as a primary alternative for parallelizing a wide range of processes on NVIDIA GPUs.

\section*{Acknowledgment}

This work was supported by the Generalitat Valenciana (GVA) through the ``Subvenciones para la contratación de personal investigador predoctoral'' under the fellowship CIACIF/2023/260.

\bibliographystyle{ieeetr}   
\bibliography{cites}  

\end{document}